\documentclass[twocolumn,showpacs,superscriptaddress,10pt]{revtex4-1}
\usepackage[colorlinks,citecolor=blue,linkcolor=red,dvipdfm]{hyperref}
\usepackage{amsmath,amssymb,graphicx}
\usepackage{times}

\begin{document}

\title{Solitons in composite linear-nonlinear moir\'{e} lattices}
\author{Liangwei Zeng}
\affiliation{School of Arts and Sciences, Guangzhou Maritime University, Guangzhou 510725, China}
\affiliation{College of Physics and Optoelectronic Engineering, Shenzhen University, Shenzhen 518060, China}

\author{Boris A. Malomed}
\affiliation{Department of Physical Electronics, School of Electrical Engineering, Faculty of Engineering,
and Center for Light-Matter Interaction, Tel Aviv University, P.O.B. 39040, Tel Aviv, Israel}
\affiliation{Instituto de Alta Investigaci\'{o}n, Universidad de Tarapac\'{a}, Casilla 7D, Arica, Chile}

\author{Dumitru Mihalache}
\affiliation{Horia Hulubei National Institute of Physics and Nuclear Engineering, 077125 Magurele, Bucharest, Romania}

\author{Jingzhen Li}
\affiliation{College of Physics and Optoelectronic Engineering, Shenzhen University, Shenzhen 518060, China}

\author{Xing Zhu}
\affiliation{School of Arts and Sciences, Guangzhou Maritime University, Guangzhou 510725, China}
\email{xingzhu@gzmtu.edu.cn}

\begin{abstract}
We produce families of two-dimensional gap solitons (GSs) maintained by moir\'{e} lattices (MLs) composed of linear and nonlinear sublattices, with the defocusing sign of the nonlinearity. Depending on the angle between the sublattices, the ML may be quasiperiodic or periodic, composed of mutually
incommensurate or commensurate sublattices, respectively (in the latter case, the inter-lattice angle corresponds to Pythagorean triples). The GSs include fundamental, quadrupole, and octupole solitons, as well as quadrupoles and octupoles carrying unitary vorticity. Stability segments of the GS families are identified by means of the linearized equation for small perturbations, and confirmed by direct simulations of perturbed evolution.
\end{abstract}

\maketitle

The creation of optical solitons \cite{Fatkh,Akhm,KA,DP,REV1,REV3,REV2,REV4,book} is one of fundamentally important topics in the broad field of nonlinear optics. In particular gap solitons (GSs) are self-trapped modes which populate bandgaps in the linear spectrum induced by a spatially periodic (lattice) linear potential, as the
result of the interplay of the bandgap spectrum with effectively self-defocusing nonlinearity \cite{Chen-Mills,GAP1,Tasgal,experimentGS1,Barash,Conti,GAP2,experimentGS2}. In addition to optics, GSs are well known as self-trapped matter-wave modes in Bose-Einstein condensates \cite{Oberthaler,Konotop,Morsch}.Various types of GSs have been demonstrated in theoretical \cite{GAP3,GAP4,GAP5} and
experimental \cite{GAP6,GAPEXP1,GAPEXP2} studies. Recently, interest has been drawn to optical \cite{Jianhua1} and matter-wave \cite{Jianhua2} GSs in moir\'{e} lattices (MLs, i.e., sets of two periodic lattices, with one rotated by a certain angle $\theta $ with respect to the other \cite{moire4}) -- in particular, due to experimental observations of two-dimensional
optical solitons of the usual (non-gap) type, maintained by the interplay of MLs and focusing nonlinearity \cite{moire1,moire2,moire3}.

It is well known that, if the value of the rotation angle $\theta $ belongs to the set of the Pythagorean triples (for instance, in the cases of $\mathrm{tan}~\theta =3/4$ or $\mathrm{tan}~\theta =5/12$), the constituent
sublattices are mutually commensurate, hence the ML keeps a periodic spatial structure. Many types of ML\ solitons have been reported recently \cite{moire5,moire6,moire7}, assuming that the underlying MLs were formed by the juxtaposition of two linear lattice potentials. On the other hand, in many
physically relevant settings spatially periodic modulation of the local nonlinearity may give rise to nonlinear lattices, characterized by the periodic \textit{pseudopotentials} \cite{REV1}.

In this Letter, we aim to consider a more specific, but experimentally feasible configuration, in which the ML structure is build as a superposition of mutually rotated linear and nonlinear lattices, with the defocusing sign of the nonlinearity, which makes it possible to construct various species of GSs.

\begin{figure}[tbp]
\begin{center}
\includegraphics[width=1\columnwidth]{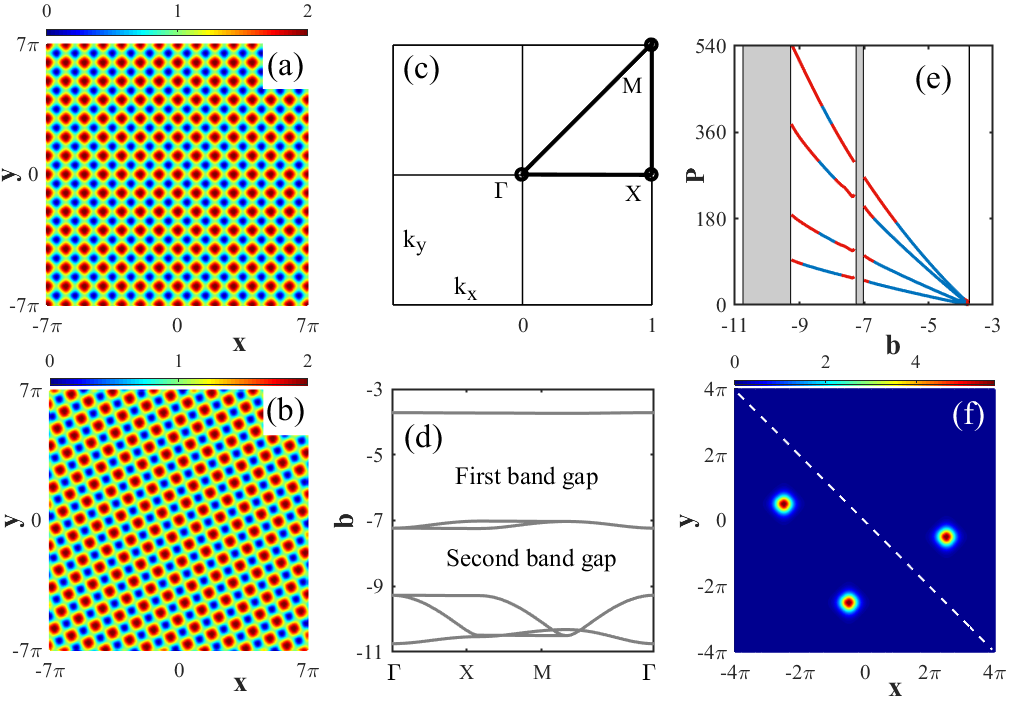}
\end{center}
\caption{(a) The linear square-lattice potential (\protect\ref{VE}) with $V_{0}=1$. (b) The distribution of local Kerr coefficient (\protect\ref{GE}), which determines the nonlinear square lattice, rotated by angle $\protect\theta =\mathrm{arctan}(5/12)$ with respect to the linear one. (c) The first reduced Brillouin zone of the linear lattice in the reciprocal lattice space. (d) The linear Bloch bandgap structure, produced by the linear potential (\protect\ref{VE}) with $V_{0}=8$. (e) Power $P$ vs. propagation constant $b$ for families of fundamental, dipole, quadrupole and octupole GSs with $m=0$ in the first two finite bandgaps (the bottom, second, third and top lines, respectively), with $V_{0}=8$ and $\protect\theta =\mathrm{arctan}(5/12)$. Blue and red segments denote stable and unstable soliton subfamilies. For the quadrupoles and octupoles with vorticity $m=1$, curves $P(b)$ and their stability structure are virtually the same as for $m=0$ (a typical soliton's power is $\sim 1$ mW in physical units).
Note that dependences $P(k)$ have no power threshold, starting from $P=0$, which is typical for GSs.
(f) The shapes of the stable fundamental and dipole solitons $|U(x,y)|$ are plotted, independently, in the areas separated by the white diagonal, for $V_{0}=8$, $b=-5.5$, and $\theta =\mathrm{arctan}(5/12)$.}
\label{fig1}
\end{figure}

The spatial-domain propagation of amplitude $E\left(x,y;z\right) $ of the optical field under the action of the composite linear-nonlinear ML is governed by the nonlinear Schr\"{o}dinger equation, written in the scaled
form:
\begin{equation}
i\frac{\partial E}{\partial z}=-\frac{1}{2}\nabla ^{2}E+V\left( x,y\right)
E+G\left( x,y\right) \left\vert E\right\vert ^{2}E.  \label{NLSE}
\end{equation}
Here, $z$ is the propagation distance, $\nabla ^{2}=\partial ^{2}/\partial x^{2}+\partial ^{2}/\partial y^{2}$ is the operator of the paraxial diffraction, and $V\left( x,y\right)$ is the potential of the linear square lattice, taken as
\begin{equation}
V\left( x,y\right) =V_{0}~\left( \mathrm{cos}^{2}x+\mathrm{cos}^{2}y\right) ,
\label{VE}
\end{equation}
with strength $V_{0}$, see it in Fig. \ref{fig1}(a). The profile of the nonlinear square lattice is defined by the spatially modulated local strength of the cubic defocusing nonlinearity [see it in Fig. \ref{fig1}(b)],
\begin{equation}
G\left( x,y\right) =\mathrm{cos}^{2}\left( x^{\prime }\right) +\mathrm{cos}
^{2}\left( y^{\prime }\right) ,  \label{GE}
\end{equation}
where ($x^{\prime },y^{\prime }$) is the coordinate system rotated by angle $\theta $: 
\begin{equation}
\left(
\begin{array}{c}
x^{\prime } \\
y^{\prime }
\end{array}
\right) =\left(
\begin{array}{cc}
~\mathrm{cos}~\theta ~ & ~-\mathrm{sin}~\theta ~ \\
~\mathrm{sin}~\theta ~ & ~\mathrm{cos}~\theta ~
\end{array}
\right) \left(
\begin{array}{c}
x \\
y
\end{array}
\right).
\label{XY2}
\end{equation}
The nonlinearity is taken with the defocusing sign, as the aim is to construct self-trapped modes of the GS type. The shape of the total effective potential can be characterized by expression
\begin{equation}
V_{\mathrm{eff}}\left( x,y\right) =\left( \mathrm{cos}^{2}x+\mathrm{cos}
^{2}y\right) +\left( \mathrm{cos}^{2}\left( x^{\prime }\right) +\mathrm{cos}
^{2}\left( y^{\prime }\right) \right),  \label{Veff}
\end{equation}
which is the sum of the linear potential (\ref{VE}) with $V_{0}=1$ and the nonlinear one with $\left\vert E\right\vert ^{2}=1$, see Eqs. (\ref{NLSE}) and (\ref{GE}). Examples of the effective potential for different values of $\theta $ are displayed below in Figs. \ref{fig2}(a1)-(f1).

Stationary solutions of Eq. (\ref{NLSE}) with real propagation constant $b$ are looked for $E\left( x,y\right) =\mathrm{exp}(ibz)U\left( x,y\right)$, with function $U$ (which is complex for vortex solitons) satisfying the equation
\begin{equation}
-bU=-\frac{1}{2}\nabla ^{2}U+V\left( x,y\right) U+G\left( x,y\right)
|U|^{2}U.
\label{NLSES}
\end{equation}
We use the modified squared-operator method \cite{Method1} to solve the stationary solutions in Eq. (\ref{NLSES}). The stationary solutions are characterized by the total power $P=\iint \left\vert U\left( x,y\right) \right\vert ^{2}dxdy$.

Typical examples of linear and nonlinear lattices (\ref{VE}) (with $V_{0}=1$) and (\ref{GE}) are displayed in Figs. \ref{fig1}(a) and (b), respectively. Further, Figs. \ref{fig1}(c) and (d) report, severally, the first reduced Brillouin zone of the linear lattice in the reciprocal lattice space, and the
respective typical Bloch bandgap structure, where one can see that both the first and second bandgaps are broad enough. These results, which are produced by the numerical solution of the linearized version of Eq. (\ref{NLSES}), are not, obviously, affected by the presence of the nonlinear lattice, being fully determined by the linear one, $V\left( x,y\right)$.

The stability analysis for the stationary solutions was performed by taking perturbed solutions as $E=[U(x,y)+p(x,y)\mathrm{exp}(\lambda z)+q^{\ast }(x,y)\mathrm{exp}(\lambda
^{\ast }z)]\mathrm{exp}(ibz)$, where $\lambda $ is the instability growth rate, $\ast $ stands for the complex conjugate, and eigenmodes $p\left( x,y\right) $ and $q\left(x,y\right) $ of small perturbations obey the linearized equations, derived by the substitution of the above expression in Eq. (\ref{NLSE}):
\begin{equation}
\begin{aligned}
i\lambda p=&-\frac{1}{2}\nabla^2p+(b+V)p+GU(2U^*p+Uq),\\
i\lambda q=&+\frac{1}{2}\nabla^2q-(b+V)q-GU^*(2Uq+U^*p).
\end{aligned}
\label{LSA}
\end{equation}
The stationary solution is stable if all eigenvalues $\lambda$ produced by the solution of Eq. (\ref{LSA}) are purely real. The eigenvalues in Eq. (\ref{LSA}) are solved by the Fourier collocation method \cite{Method2}.

\begin{figure*}[tbp]
\begin{center}
\includegraphics[width=2\columnwidth]{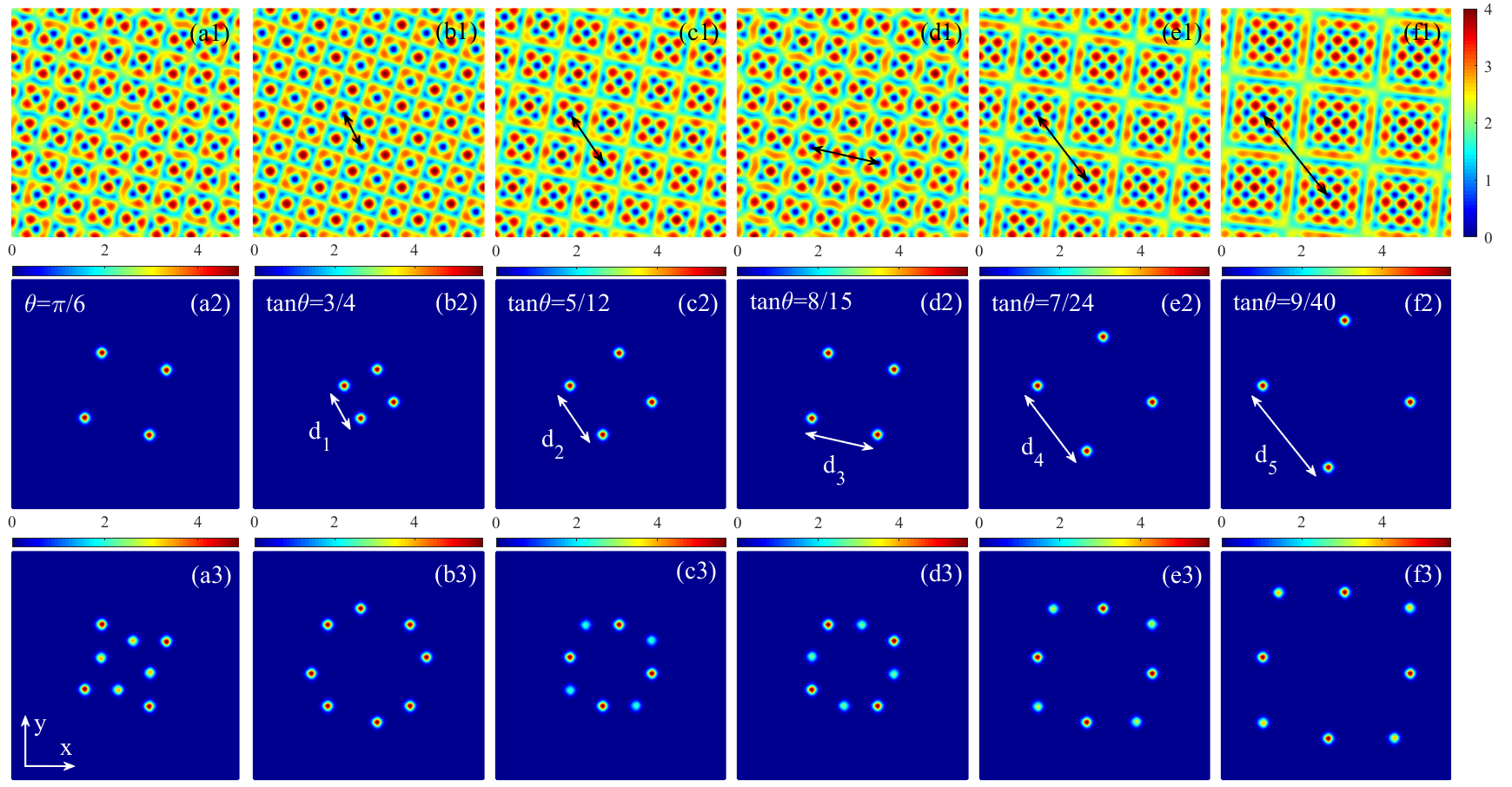}
\end{center}
\caption{The top row: the quasiperiodic effective potential (\protect\ref{Veff}) for $\protect\theta =\protect\pi /6$ in panel (a1), and periodic potentials for values of $\protect\theta $ from Eq. (\protect\ref{theta}) in panels (b1)-(f1). The middle and bottom rows: shapes ($|U|$) of the quadrupole and octupole solitons, respectively, as produced by Eq. (\protect\ref{NLSES}) with $b=-5.5$ and the MLs composed of linear and nonlinear lattice potentials defined as per Eqs. (\protect\ref{VE})-(\protect\ref{XY2}). Black and white arrows in the top and middle panels denote periods of effective
potentials. Solitons in all panels are stable. All panels display domains $\left\vert x,y\right\vert \leq 7\protect\pi$.
}
\label{fig2}
\end{figure*}

The families of such solitons, including fundamental, dipole, quadrupole and octupole ones, produced by the numerical solution of Eq. (\ref{NLSES}), are summarized in Fig. \ref{fig1}(e) by means of curves showing the respective power $P$ vs. the propagation constant $b$, blue and red segments representing stable and unstable subfamilies, respectively (the curves for the quadrupole and octupole solitons with vorticity $m=1$ are almost the same as the ones for $m=0$). Note that the $P(k)$ curves satisfy the \textit{anti-Vakhitov-Kolokolov criterion}, $dP/db<0$, which is a necessary (but generally, not sufficient) condition for the stability of solitons (such as GSs) maintained by nonlinearity with the self-defocusing sign \cite{AVK} (the Vakhitov-Kolokolov criterion proper, has the opposite form, $dP/db>0$, providing a necessary stability condition for solitons in systems
with self-focusing nonlinearities \cite{VK,Berge}). An example of stable fundamental and dipole solitons (divided by a white dashed line) is presented in Fig. \ref{fig1}(f). Here the center points of such fundamental and dipole soliton are ($x=2.5\pi$,$y=-0.5\pi$) and ($x=-2.5\pi,y=0.5\pi$; $x=-0.5\pi,y=-2.5\pi$), respectively.

\begin{figure}[htbp]
\begin{center}
\includegraphics[width=1\columnwidth]{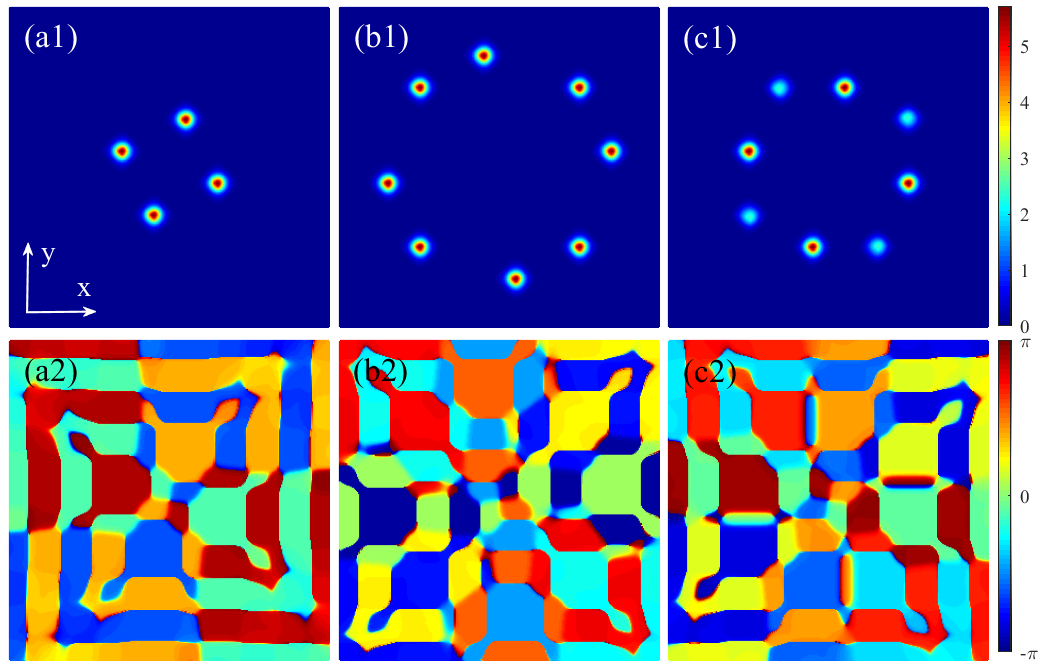}
\end{center}
\caption{The top and bottom rows display the power ($|U|$) and phase structures of the vortex solitons with propagation constant $b=-5.5$ and winding number $m=1$: (a1,a2) a quadrupole soliton with $\protect\theta =\mathrm{\arctan}
(3/4)$; (b1,b2) and (c1,c2): octupole solitons with the ML angles $\protect\theta =\mathrm{\arctan }(3/4)$ and $\mathrm{\arctan }(5/12)$, respectively. Solitons in all panels are stable. All panels display domains $\left\vert x,y\right\vert \leq 5\protect\pi$.}
\label{fig3}
\end{figure}

Examples of multipole solitons with four and eight peaks are presented in Fig. \ref{fig2}, for various effective linear-nonlinear potentials, which are plotted in the top row of the figure, as per Eq. (\ref{Veff}). Density profiles of the four-peak (quadrupole) and eight-pole (octupole) solitons are displayed, respectively, in the middle and bottom rows of Fig. \ref{fig2}.

Column (a) in Fig. \ref{fig2} reports the results for $\theta =\pi /6$, which, with irrational $\cos (\pi /6)=\allowbreak \sqrt{3}/2$, makes the effective (composite) potential (\ref{Veff}) quasiperiodic. The other columns, (b) -- (f), represent various values of $\theta $ which correspond
to the Pythagorean triples, namely,
\begin{gather}
\theta =\arctan (3/4)~(\mathrm{b}),~\theta =\mathrm{\arctan }(5/12)~(\mathrm{
c}),~\theta =\mathrm{\arctan }(8/15)~(\mathrm{d}),~  \notag \\
\theta =\mathrm{\arctan }(7/24)~(\mathrm{e}),\arctan \theta =\mathrm{\arctan
}(9/40)~(\mathrm{f}).
\label{theta}
\end{gather}
In these cases, the composite potential (\ref{Veff}) is periodic, with the respective periods shown by black arrows in panels (b1)-(f1) of Fig, \ref{fig2}. Further, white arrows in panels (b2)-(f2) show that the same periods determine positions of local peaks of the quadrupole solitons and distances $\mathrm{d}_{m}$ ($m=1,...,5$) between them. For instance, in Fig. \ref{fig2}(b1), two peaks connected by the white arrows are located at points $\left(x,y\right) =(-1.5\pi ,0.5\pi )$ and $(-0.5\pi ,-1.5\pi )$, hence the distance between them is $\mathrm{d}_{1}\approx 7.02$. In the other panels, the periods are $\mathrm{d}_{2}\approx 11.33$ in (c2), $\mathrm{d}_{3}\approx 12.95$ in (d2), $\mathrm{d}_{4}\approx 15.71$ in (e2), and $\mathrm{d}_{5}\approx 20.12$ in (f2). These values demonstrate that the period of the ML composed of commensurable linear and nonlinear lattice potentials naturally increases with the growth of the order of the corresponding Pythagorean triples.

The quasiperiodic character of the composite ML potential (\ref{Veff}) in the case of $\theta =\pi /6$ leads to a difference of the peak values of the local power between two quartets of local maxima which constitute the octupole structure in panel (a3) of Fig. \ref{fig2}. Note also that the
effective ML potential (\ref{Veff}) provides\ only an approximate description of the setting, as the actual nonlinear potential depends on the distribution of the local power in the stationary state. In fact, the description is quite accurate for $\theta =\arctan (3/4)$, therefore the
corresponding octupole soliton, displayed in Fig. \ref{fig2}(b3), features equal powers at eight peaks of the pattern. On the other hand, for other values of the rotation angle, \textit{viz}., $\theta =\mathrm{\arctan }(5/12)$, $\theta =\mathrm{\arctan }(8/15)$, $\theta =\mathrm{\arctan }(7/24)$, and $\arctan \theta =\mathrm{\arctan }(9/40)$ in Eq. (\ref{theta}), it is seen in Figs. \ref{fig2}(c3)-(f3) that the octupole actually splits into a juxtaposition of two quadrupoles with different peak powers, hence the effective potential does not provide a sufficiently accurate prediction in those cases.

Typical examples of power and phase profiles of quadrupole and octupole solitons with embedded topological charge (winding number) $m=1$ are presented in Fig. \ref{fig3}. In particular, similar to the above-mentioned difference between the zero-vorticity octupoles displayed in Figs. \ref{fig2}(b3) and (c3) for $\theta =\arctan (3/4)$ and $\arctan \left( 5/12\right) $,
Figs. \ref{fig3}(b1) and (c1) demonstrate that the vortex octupole corresponding to the former value of $\theta$ features equal powers of all peaks, while one corresponding to the latter value of $\theta $ splits into a juxtaposition of two quadrupoles with different peak powers.

\begin{figure}[tbp]
\begin{center}
\includegraphics[width=1\columnwidth]{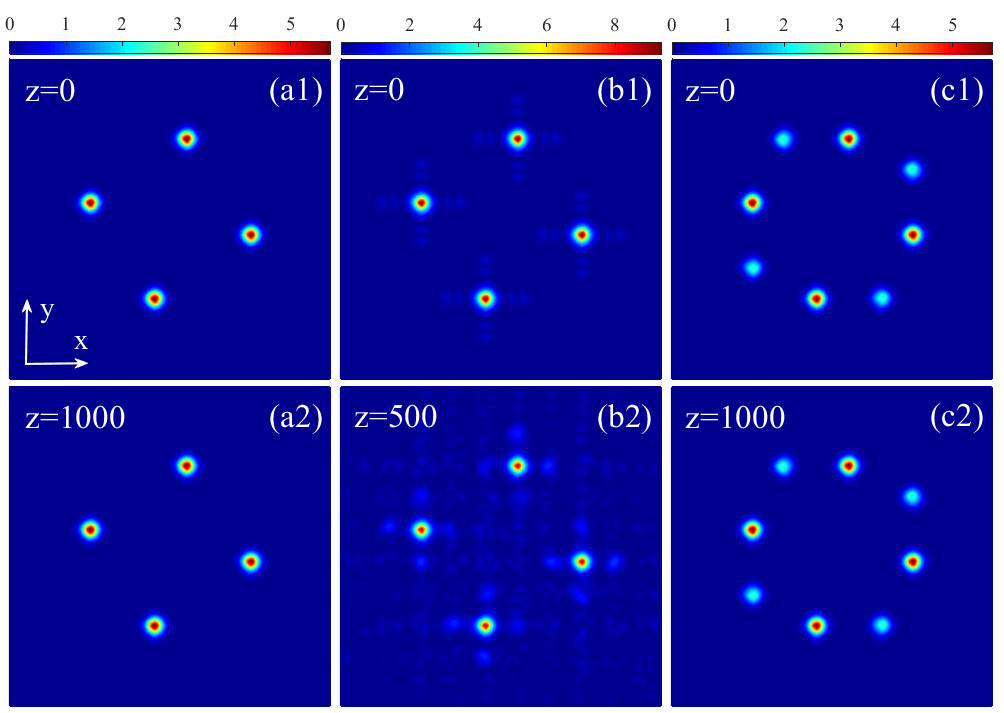}
\end{center}
\caption{The perturbed propagations of solitons in ML with the rotation
angle $\protect\theta =\mathrm{\arctan }(5/12)$. (a1,a2) A stable quadrupole
soliton with zero vorticity, $m=0$, and propagation constant $b=-5.5$. (b1,
b2) An unstable quadrupole soliton with $m=0$ and $b=-7.65$. (c1,c2) A stable octupole soliton with $m=1$ and $b=-5.5$. All
panels display domains $\left\vert x,y\right\vert \leq 5\protect\pi $ .}
\label{fig4}
\end{figure}

Finally, examples of simulations of the propagation of quadrupole and octupole solitons, both zero-vorticity and vortical ones, perturbed by random complex noise with the relative amplitude $\sim 1\%$, which corroborate the prediction for the (in)stability produced by the computation of eigenvalues $\lambda $ in the framework of linearized equations (\ref{LSA}), are presented in Fig. \ref{fig4}. The top and bottom rows in the figure show the input and output of the perturbed evolution, corresponding to the propagation distance $z=1000$, which may be estimated as $\sim 4$ diffraction lengths (longer simulations do not change the outcome of the apparently stable evolution). The left and right columns
demonstrate, severally, the stable propagation of a zero-vorticity quadrupole and octupole with embedded vorticity $m=1$, while unstable propagation of an unstable zero-vorticity quadrupole soliton is displayed in the middle column. The unstable solitons suffer full decay in the course of longer evolution.

To conclude, we have introduced MLs (moir\'{e} lattices) composed of linear and nonlinear sublattices, with mutual rotation angle $\theta $, and the defocusing sign of the nonlinearity. By means of numerical methods and qualitative
consideration of the effective composite potential, we have constructed families of fundamental (single-peak), dipole (two-peak), quadrupole (four-peak), and octupole (eight-peak) GSs (gap solitons) with zero-vorticity, as well as quadrupole
and octupole GSs with embedded vorticity $m=1$. Stability segments are identified in these families. Evolution of unstable GSs is briefly considered by means of direct simulations. Considered are both the systems of mutually incommensurate linear and nonlinear sublattices (in particular,
ones with $\theta =\pi /6$) and commensurate pairs of sublattices, with $\theta $ corresponding to Pythagorean triples (as given by Eq. (\ref{theta})). The size of the quadrupole and octupole solitons increases with the growth of the order of the triples. As an extension of the work, it may be interesting to consider MLs formed by two nonlinear sublattices, with the same or opposite signs of the intrinsic
nonlinearities.
\\\\
\noindent\textbf{Funding.} National Natural Science Foundation of China (62205224; 61827185); Guangdong Basic and Applied Basic Research Foundation (2023A1515010865); Start-up Foundation for Talents of Guangzhou Jiaotong University (K42022076, K42022095); Israel Science Foundation (1695/22).
\\
\noindent\textbf{Disclosures.} The authors declare no conflicts of interest.
\\
\noindent\textbf{Data availability.} Data underlying the results presented in this paper may be obtained from the authors upon request.

\end{document}